\documentclass[journal,twoside]{IEEEtran}

\usepackage{cite}
\usepackage{amsmath,amssymb}
\usepackage{graphicx}
\usepackage{array}
\usepackage{booktabs}
\usepackage{siunitx}
\usepackage{textcomp}
\usepackage{url}
\usepackage{balance}

\usepackage{makecell}
\usepackage{xcolor}
\newcommand{\revThreeColor}{black}
\DeclareRobustCommand{\revThree}[1]{\textcolor{\revThreeColor}{#1}}

\graphicspath{{images/}}
\DeclareGraphicsExtensions{.pdf,.png,.jpg}

\DeclareSIUnit{\ppm}{ppm}
\DeclareSIUnit{\torr}{Torr}

\begin{document}

\title{
Thin-Film AlN Microbolometer for
Very-Long-Wave Infrared Detection
}

\author{
Vivek~Tallavajhula,
Zarko~Sakotic,
Ian~Anderson,
Yinan~Wang,
Daniel~Wasserman,
and~Ruochen~Lu%
\thanks{
This work has been submitted to the IEEE for possible publication. Copyright may be transferred without notice, after which this version may no longer be accessible.
This work was supported in part by the Defense Advanced Research Projects Agency Optomechanical Thermal Imaging (OpTIm) program.  Z.S. and D.W. acknowledge support from the University of Dayton Research Institute via Air Force (Grant No. FA8651-20-D-0003).
(Corresponding author: Vivek Tallavajhula.)
}%
\thanks{
The authors are with the Department of Electrical and Computer
Engineering, The University of Texas at Austin, Austin, TX 78712 USA
(e-mail: vivek.tallav@utexas.edu).
}%
}

\markboth{
}%
{
Tallavajhula \MakeLowercase{\textit{et al.}}:
Thin-Film AlN Microbolometer
}

\maketitle

\begin{abstract}
We demonstrate a suspended thin-film aluminum nitride (AlN)
microbolometer for narrowband very-long-wave infrared detection.
The device uses a \SI{100}{\nano\meter}-thick AlN membrane
suspended above a Pt back reflector by a \SI{1}{\micro\meter}
air gap. Resonant absorption near \SI{15.4}{\micro\meter} is set
by the AlN transverse optical phonon and is strengthened by
suspension above the reflector. A periodic perforation pattern
reduces membrane thermal mass, facilitates release, and tunes the
effective optical thickness without requiring a thinner deposited film.
Fourier-transform infrared (FTIR) reflection measurements on the released bolometer show approximately
95\% peak cavity absorptance near \SI{15.4}{\micro\meter} with a
\SI{0.772}{\micro\meter} FWHM. DC resistance measurements under tunable
infrared illumination verify bolometric operation, with a peak normalized
resistive responsivity of \SI{252}{\ppm\per\milli\watt} at
\SI{15.48}{\micro\meter}. This platform can enable compact
wavelength-selective thermal detectors for multispectral imaging,
on-chip infrared spectroscopy, and chemical sensing.
\end{abstract}


\section{Introduction}

\IEEEPARstart{I}{nfrared} (IR) detection supports various
applications in defense, industrial monitoring, medical diagnostics,
and environmental sensing
\cite{
saleemInfraredPhotodetectorsRecent2024,
kozuchSurfaceenhancedInfraredAbsorption2023
}.
The long-wave infrared (LWIR, \SIrange{8}{12}{\micro\meter}) and
very-long-wave infrared (VLWIR, \SIrange{12}{30}{\micro\meter})
regimes, in particular, are widely used for thermal imaging and
remote sensing, a result of their position as the spectral home for
thermal emission from objects at or below room temperature
\cite{talghaderSpectralSelectivityInfrared2012}.
Absorber regions within LWIR detectors enable efficient conversion
of optical power into either thermal or electrical energy
\cite{
talghaderSpectralSelectivityInfrared2012,
nordinUltrathinPlasmonicDetectors2021
}.
The latter approach, photon detectors most often using semiconductor
materials, typically offers higher-bandwidth detection but requires
cryogenic cooling and epitaxial growth
\cite{wangHighspeedLongwaveInfrared2024}.
The former, generally described as microbolometers, sense infrared
radiation through a temperature-induced change in an electrical
signal. While bolometric detection cannot match the response speed
of photon detectors, it does enable uncooled operation and low-cost,
large-area VLWIR imaging arrays. Realizing such absorbers in
thin-film platforms is nontrivial, as strong absorption must be
achieved within subwavelength thicknesses while maintaining
compatibility with suspended, thermally isolated structures. This
challenge is especially relevant for compact, uncooled platforms,
where absorber performance requires strong optical coupling and
thermal confinement
\cite{
abdullahUncooledTwomicrobolometerStack2023,
alkorjiaMetasurfaceEnabledUncooled2023
}.

The traditional approach to achieving strong absorption in the
VLWIR range broadly relies on a thin-film absorption approach
\cite{woltersdorffBerOptischenKonstanten1934}, where lossy films such
as titanium, platinum
\cite{martiniUncooledThermalInfrared2025}, titanium nitride
\cite{huoBroadbandAbsorptionBased2021}, and more recently
two-dimensional materials such as transition-metal carbides,
nitrides, carbonitrides, and atomically thin Au films
\cite{
zhaoUltrathinMXeneAssemblies2023,
luhmannUltrathin2Nm2020
}
are used, but these are usually ultrabroadband in character and offer
no selectivity, while allowing absorption only up to the 50\%
thin-film absorption limit. On the other hand, selective absorption
in this range can be achieved with the classic metamaterial
approach, structuring metals and other conductive layers, usually
implemented in metal--insulator--metal architectures, thereby
enabling frequency-selective perfect absorption
\cite{
talghaderSpectralSelectivityInfrared2012,
jungWavelengthSelectiveInfraredMetasurface2015
}.
However, the selectivity remains relatively poor due to the lossy
nature of metals in this range.

An alternative, somewhat underutilized, approach to highly selective
absorption in this range is to leverage intrinsic material
resonances, such as those in polar materials like hexagonal boron
nitride, silicon carbide, and aluminum nitride
\cite{
caldwellLowlossInfraredTerahertz2014a,
foteinopoulouPhononpolaritonicsEnablingPowerful2018
},
whose optical phonon vibrations lie across the VLWIR range.
Conventional wisdom suggests that having more absorptive material
will lead to more absorption; however, recent theoretical work shows
that such phononic layers can absorb 100\% of incident light in
thicknesses thousands of times smaller than their phonon wavelengths
(\(<\SI{10}{\nano\meter}\))
\cite{sakoticPerfectAbsorptionUltimate2023}, which lends itself
perfectly to the task at hand, i.e., construction of low-volume,
low-mass, selective absorbers for bolometric applications. While
using phonon polaritons directly for thermal detection has been
proposed previously
\cite{gubbinSurfacePhononPolaritons2022}, to our knowledge, no such device has been demonstrated to date.

In this work, we demonstrate a suspended thin-film aluminum nitride
microbolometer that combines a platinum back reflector, a
molybdenum-defined air cavity, and a perforated AlN membrane with
integrated resistive readout, where resonant optical absorption in
the suspended membrane produces a measurable change in resistance.

To connect the optical and electrical characterization, the same
released bolometer is examined in two complementary steps. We
first characterize the intrinsic spectral response by FTIR spectroscopy
performed directly on the suspended absorber region and compare the result
with optical simulations. We then illuminate the device with a tunable
infrared source and record the thermoresistive response. Thus, the FTIR
measurement establishes the spectral absorber characteristics, while the
optical measurement with electrical readout verifies bolometric operation.

\section{Device Architecture and Fabrication}

Aluminum nitride is selected as the absorber material for its high
mechanical stiffness, with Young's modulus approximately
\SI{330}{\giga\pascal}, which helps keep the membrane suspended
across a \SI{1}{\micro\meter} air gap after release
\cite{
iqbalReactiveSputteringAluminum2018,
mastrangeloAdhesionrelatedFailureMechanisms1997
},
and for its strong transverse optical (TO) phonon resonance near
\SI{15.0}{\micro\meter} at the long-wavelength end of its VLWIR
Reststrahlen band. Additionally, as a CMOS-compatible material,
thin-film AlN can be sputter-deposited with high crystalline quality
on a variety of substrates
\cite{
liuPiezoelectricThinFilms2025,
iqbalReactiveSputteringAluminum2018,
chenEnhancementCAxisOriented2021
}.

By integrating a subwavelength AlN membrane into a suspended-cavity
architecture, we enable critical coupling between incident radiation
and the TO phonon mode. The geometry of the perforated membrane
addresses mechanical reliability issues in thin sputtered AlN films,
which are susceptible to high compressive stress and
stiction-induced collapse during release at thicknesses below
\SI{200}{\nano\meter}
\cite{
lundhResidualStressAnalysis2021,
iqbalReactiveSputteringAluminum2018,
tremlHighResolutionDetermination2016
}.
The perforations reduce thermal mass \(C_{\mathrm{th}}\), thereby
improving response time \(\tau\) without further thinning the film;
they also serve to engineer the film's absorption properties, which
we discuss further below. Mo serves as both the sacrificial spacer
and the seed layer for highly (002)-oriented AlN sputtering
\cite{felmetsgerCrystalOrientationStress2008}, and is selectively
removable via XeF$_2$ dry etching. The underlying Pt layer acts as a broadband IR back reflector
and suppresses transmission through the device. This can be verified
by per-layer absorption simulations (see Supplementary Fig.~2), which
demonstrate that absorption in AlN dominates and Pt absorption is small
(\(<7\%\)) within the cavity. We can therefore approximate the measured
quantity
\begin{equation}
A \approx 1-R,
\label{eq:A_from_R}
\end{equation}

as the total absorptance of the AlN/air-gap/Pt cavity, where \(A\) and
\(R\) are absorptance and reflectance, respectively
\cite{novotny_principles_2012}. The sapphire substrate provides a
low-roughness foundation for the stack.

\begin{figure}[!t]
\centering
\includegraphics[width=\columnwidth]{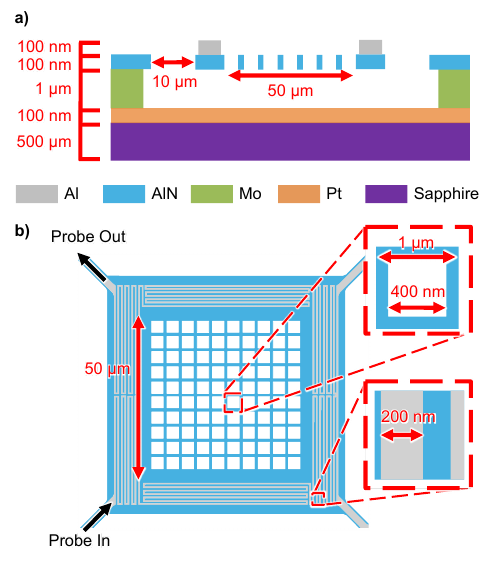}
\caption{
Device geometry of the suspended AlN microbolometer.
(a) Cross-sectional schematic of the suspended AlN absorber above a
Pt back reflector, with a Mo spacer defining the air gap.
(b) Top-view layout of the perforated membrane, routed metal traces,
and corner tethers.
}
\label{fig:device}
\end{figure}

The detector architecture is shown in Fig.~\ref{fig:device}. A
\SI{100}{\nano\meter} AlN membrane with a patterned top Al layer is
suspended above a Pt back reflector by a \SI{1}{\micro\meter} air
cavity. The AlN membrane provides the resonant optical absorption
near the TO phonon. The patterned Al trace provides the thermoresistive
readout of the resulting temperature rise. Corner tethers provide mechanical support and a limited
thermal conduction path. A square array of
\SI{400}{\nano\meter}$\times$\SI{400}{\nano\meter} perforations is
defined in the suspended region, which also serves as access paths
for XeF$_2$ during release.

\begin{figure}[!t]
\centering
\includegraphics[width=\columnwidth]{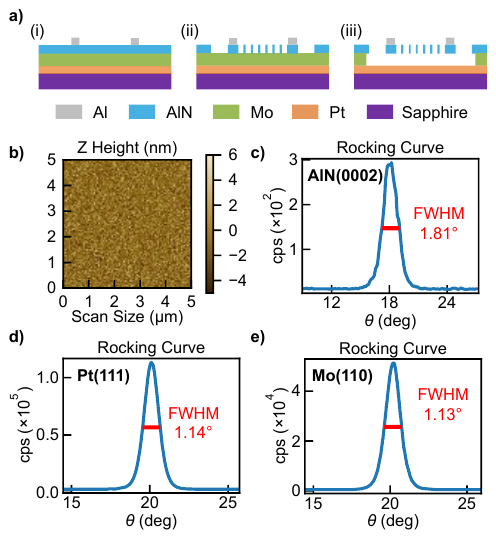}
\caption{
Fabrication sequence and material characterization of the suspended
VLWIR absorber.
(a) Fabrication process flow.
(b) AFM surface topography of the deposited AlN film.
(c)--(e) X-ray diffraction rocking curves of AlN(0002), Pt(111),
and Mo(110), respectively, with extracted FWHM values of
\(1.81^\circ\), \(1.14^\circ\), and \(1.13^\circ\).
}
\label{fig:fabrication}
\end{figure}

The device is fabricated using a multilayer stack deposited on a
2-inch sapphire substrate with an Evatec sputtering system. The
sequence begins with the deposition of a
\SI{100}{\nano\meter} Pt back reflector, followed by a
\SI{1}{\micro\meter} sacrificial Mo spacer. Both layers demonstrate
excellent crystalline uniformity as evidenced by FWHM measurements
of \(1.14^\circ\) and \(1.13^\circ\) for the Pt and Mo films,
respectively [Fig.~\ref{fig:fabrication}(d),(e)]. A
\SI{100}{\nano\meter} AlN device layer is then deposited via pulsed
DC reactive magnetron sputtering. AFM and XRD characterization of
the as-deposited stack, as shown in
Fig.~\ref{fig:fabrication}(b),(c), confirms high crystalline quality,
with a measured FWHM of \(1.81^\circ\) for the AlN(0002) peak and a
surface roughness \(R_a\) of \SI{1.12}{\nano\meter}. The measured
AlN(0002) FWHM of \(1.81^\circ\) confirms sufficient \(c\)-axis
orientation to minimize inhomogeneous phonon broadening and preserve
a sharp resonant response
\cite{chenEnhancementCAxisOriented2021}.

After stack deposition, two electron-beam lithography steps define
(i) the Al resistive traces and probing pads via evaporation and
lift-off and
(ii) the \SI{400}{\nano\meter} perforations and
\SI{10}{\micro\meter} etch windows, transferred through AlN and Mo by
ion milling [Fig.~\ref{fig:fabrication}(a)]. The sacrificial Mo is
then removed by timed XeF$_2$ gas-phase etching, where the
perforation array provides a uniform etch front, yielding a fully
suspended membrane while preserving Mo support beneath the contact
pads.

\section{Optical Response and Absorber Design}

\subsection{Permittivity and Reflection Modeling}

The reflection coefficient is obtained using the ABCD (transfer)
matrix method. Using transmission-line theory, each optical layer
can be represented by its respective ABCD (transfer) matrix
\begin{equation}
\mathbf{M}_n
=
\begin{pmatrix}
\cos(k_n d_n)
&
-i Z_n \sin(k_n d_n)
\\
-i\sin(k_n d_n)/Z_n
&
\cos(k_n d_n)
\end{pmatrix},
\label{eq:layer_matrix}
\end{equation}
where \(k_n\), \(d_n\), and \(Z_n\) are the perpendicular wavevector
component, thickness, and wave impedance of each layer. These are
calculated as
\begin{equation}
\begin{aligned}
k_n
&=
\frac{2\pi\sqrt{\epsilon_n}}{\lambda}
\cos\theta_n,
\\
Z_n^{TM}
&=
\frac{Z_0}{\sqrt{\epsilon_n}}
\cos\theta_n,
\\
Z_n^{TE}
&=
\frac{Z_0}
{\sqrt{\epsilon_n}\cos\theta_n},
\end{aligned}
\label{eq:wavevector_impedance}
\end{equation}
where \(Z_0\) is the free-space impedance and \(\theta_n\) is the
angle of refraction in each layer. The total matrix is then:
\begin{equation}
\mathbf{M}_T
=
\mathbf{M}_{AlN}
\mathbf{M}_{gap}
\mathbf{M}_{Pt}
=
\begin{pmatrix}
A & B\\
C & D
\end{pmatrix}.
\label{eq:total_matrix}
\end{equation}

From there, the reflection coefficient is calculated as
\begin{equation}
r
=
\frac{
A + B/Z_0 - CZ_0 - D
}{
A + B/Z_0 + CZ_0 + D
}.
\label{eq:reflection_coefficient}
\end{equation}

\begin{figure}[!t]
\centering
\includegraphics[width=\columnwidth]{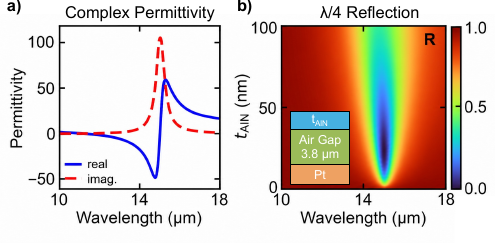}
\caption{
Optical properties and ultimate-thickness cavity design for AlN.
(a) Frequency-dependent real and imaginary components of the AlN
permittivity calculated using the single-Lorentzian model.
(b) Reflection as a function of wavelength and AlN thickness for a
cavity close to the quarter-wave condition with a
\SI{3.8}{\micro\meter} air gap. In this case, the absorption remains
above 90\% at thicknesses near \SI{10}{\nano\meter}.
}
\label{fig:ultimate_thickness}
\end{figure}

The permittivity of AlN is modeled with a single Lorentzian written
in wavenumber notation,
\begin{equation}
\epsilon(\tilde{\nu})
=
\epsilon_\infty
\left(
1+
\frac{
\tilde{\nu}_{\mathrm{LO}}^2-\tilde{\nu}_{\mathrm{TO}}^2
}{
\tilde{\nu}_{\mathrm{TO}}^2-\tilde{\nu}^2-i\tilde{\gamma}\tilde{\nu}
}
\right),
\label{eq:lorentz}
\end{equation}
where
$\tilde{\nu}_{LO}=885~\mathrm{cm}^{-1}$,
$\tilde{\nu}_{TO}=665~\mathrm{cm}^{-1}$,
$\tilde{\gamma}=22.5~\mathrm{cm}^{-1}$, and
$\epsilon_\infty=4.6$.
The complex relative permittivity is plotted in Fig.~\ref{fig:ultimate_thickness}(a).

\subsection{Quarter-Wave Cavity}

We use the same model to examine the ultimate absorber-thickness
limit obtainable with an appropriately adjusted cavity.

Previously, it has been shown that for a resonantly absorbing layer
such as AlN placed over a quarter-wave reflecting cavity, perfect
absorption can be achieved when the thickness of the layer is exactly
\cite{sakoticPerfectAbsorptionUltimate2023}
\begin{equation}
t_w
=
\frac{\lambda_{\mathrm{TO}}}
{2\pi\epsilon''},
\label{eq:woltersdorff}
\end{equation}
where \(\lambda_{\mathrm{TO}}\) represents the resonant TO wavelength
and \(\epsilon''\) is the imaginary permittivity value at its resonant
peak. This simple formula represents the so-called Woltersdorff
thickness
\cite{woltersdorffBerOptischenKonstanten1934}, which was generalized
to resonant material response
\cite{sakoticPerfectAbsorptionUltimate2023} and previously confirmed
in different material systems
\cite{sakoticMidInfraredPerfectAbsorption2024}.

This predicts 100\% absorption in an AlN layer only
\SI{23}{\nano\meter} thick when placed over
\(\lambda_{\mathrm{TO}}/4=\SI{3.8}{\micro\meter}\) air gap, assuming
the same optical quality as measured. The corresponding AlN
permittivity and quarter-wave thickness sweep are shown in
Fig.~\ref{fig:ultimate_thickness}. For this appropriately adjusted
air gap, absorption remains above 90\% at thicknesses near
\SI{10}{\nano\meter}. However, due to the smaller air gap in our
system, \SI{1}{\micro\meter}, the perfect-absorption condition is
predicted to occur at a thickness of \SI{120}{\nano\meter}, while the
absorption is predicted to remain above 90\% down to
\SI{50}{\nano\meter}.

\begin{figure*}[!t]
\centering
\includegraphics[width=0.99\textwidth]{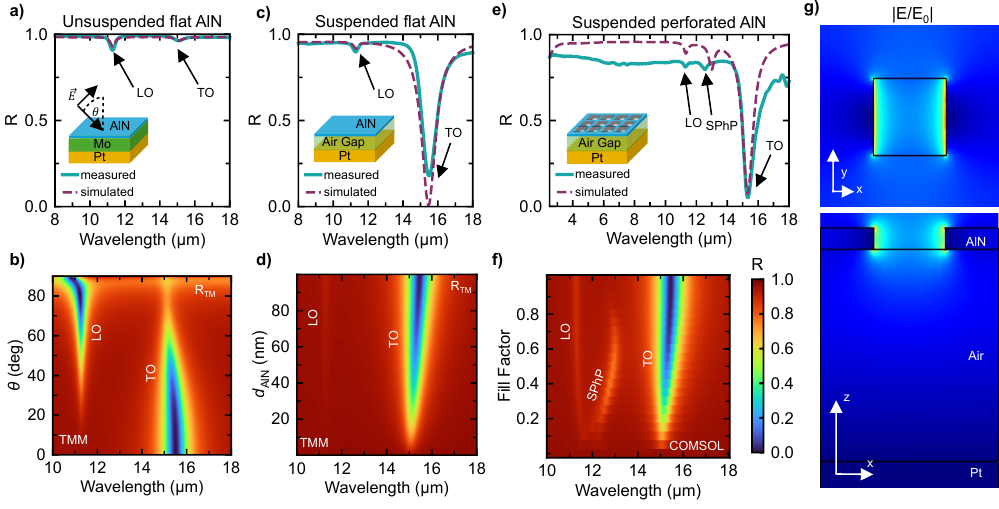}
\caption{
(a) Measured and simulated reflection of the AlN thin-film cavity in
the unsuspended and (c) suspended case.
(b) Simulated reflection of the suspended AlN thin-film cavity as a
function of wavelength and incident angle for TM-polarized light and
(d) as a function of wavelength and film thickness.
(e) Measured and simulated reflection of the suspended, perforated
AlN membrane in the extended spectral range, with three main phonon
features designated.
(f) Numerically simulated reflection of the perforated, suspended AlN
membrane cavity as a function of wavelength and perforation size,
i.e., membrane fill fraction.
(g) Numerical simulation of the electric fields at the TO resonance.
}
\label{fig:optical}
\end{figure*}

\subsection{Angular and Thickness Dependence}

To understand the optical properties of the AlN layer in the VLWIR, we
begin by analyzing Fourier-transform infrared (FTIR) reflection spectroscopy
of the samples, as shown in Fig.~\ref{fig:optical}(a),(c).
The FTIR measurements are performed on the actual bolometer, without
the use of passive test structures. In order to avoid interference from the
surrounding reflective metal used for resistive readout, we focus the FTIR
beam onto only the \SI{50}{\micro\meter} $\times$ \SI{50}{\micro\meter}
perforated absorber area. Due to the
high reflectivity of Mo and Pt layers, the transmission through the sample
is fully suppressed in both the suspended and unsuspended cases. Initially, we analyze an
unsuspended structure, i.e., AlN backed by \SI{1}{\micro\meter} of Mo
and \SI{100}{\nano\meter} of Pt. As expected, the system reflects
light almost perfectly, except for weak features at the LO and TO
phonon frequencies, i.e., around \SI{11.3}{\micro\meter} and
\SI{15.4}{\micro\meter}.

Simulation of this simple structure using the transfer-matrix method
shows that we can accurately reproduce the measured features,
thereby confirming the high optical quality of AlN. Next, we move to
the suspended structure, i.e., a structure in which the Mo
sacrificial layer has been removed, leaving an air gap, which,
together with the Pt mirror and the AlN top layer, forms a simple
Fabry--P\'erot cavity. A simple transmission-line model of the cavity
can accurately predict its optical properties and guide the absorber
design.

To illuminate the origin and behavior of these modes, we also
simulate the reflection of the cavity system as a function of the
incidence angle and AlN thickness in
Fig.~\ref{fig:optical}(b),(d). The TO feature remains quite robust
across a large range of incident angles and polarizations, ensuring
absorption from a broad angular range. The LO mode, also known as the
epsilon-near-zero or Berreman mode, requires large-angle,
TM-polarized light for maximum coupling
\cite{nordinMidinfraredEpsilonnearzeroModes2017}, which explains the
weak feature observed in our experimental results. Indeed, the
simulations indicate that near-perfect absorption is possible at
angles above \(70^\circ\) for TM-polarized light.

While this configuration does not necessarily benefit from
absorption or detection at the LO frequency, the simulated strong
absorption is aligned spectrally with the room-temperature thermal
emission peak at \SIrange{10}{11}{\micro\meter}. The simulated LO
resonance may provide an additional radiative-relaxation pathway under
suitable angle and polarization, but this pathway is not isolated
experimentally in the present work \cite{wareDecouplingAbsorptionRadiative2023}.
The thickness-dependent simulation shows that strong absorption at the
TO mode is preserved down to \SI{50}{\nano\meter}. This has remarkable
implications for the potential speed and sensitivity of detection, since
the massively decreased absorber volume does not come at a price of reduced
efficiency.

As predicted, the experimentally measured reflection shows very strong,
selective absorption at the TO frequency, Fig.~\ref{fig:optical}(c),
reaching over 80\%, though somewhat smaller than the simulation, which
predicts near-perfect absorption at \SI{15.4}{\micro\meter}. The
measured flat suspended cavity reaches greater than 80\% peak absorptance
near the TO mode with \(Q\approx17\). Together with the
\SI{100}{\nano\meter} AlN thickness, this demonstrates strong narrowband
absorption in an ultrathin cavity-coupled polar film. For comparison, the
present absorber has \(t/\lambda_{\mathrm{TO}}\approx1/150\), smaller than
prior cavity-coupled polar absorbers
\cite{lingNearUnityLightMatter2024,abedinidereshgiLithographyfreeIRPolarization2020}.
The absorption in the LO mode remains low due to low coupling efficiency
under the experimental excitation conditions.

\subsection{Perforated Absorber Response}

However, growing high-quality AlN at such low thicknesses is
challenging, as the initial layers during deposition tend to be of
poor quality
\cite{beliaev2021thickness} and are likely to lose the optical
quality required for efficient absorption at these thicknesses.
Instead, absorber volume and thermal mass can be reduced via
subwavelength perforations rather than using thinner layers
\cite{sakoticMidInfraredPerfectAbsorption2024}, without any cost to
optical quality; it was recently shown that optical properties of
atomically thin layers can be mimicked by perforating a slightly
thicker film on a subwavelength scale
\cite{sakoticMidInfraredPerfectAbsorption2024}.
This approach can improve the optical, thermal, and mechanical
stability of the absorber layer, as mentioned previously.

The perforations provide distributed XeF$_2$ access, reduce
absorber volume and heat capacity, and lower the effective AlN fill factor.
In the TO band, changing fill factor produces a similar optical shift to
reducing the AlN thickness, providing a tuning mechanism without a thinner
deposited film.

Due to the ultrathin nature of the layer, the optical response scales
equally with both fill factor and thickness; numerical simulations
of the film perforated with holes at a \SI{1}{\micro\meter} period
shown in Fig.~\ref{fig:optical}(f) show that fill-factor-dependent
optical properties of the structure very closely resemble thickness
reduction in the TO band [Fig.~\ref{fig:optical}(d)]. Additionally,
the subwavelength and periodic nature of the perforations gives rise
to surface-phonon-polariton (SPhP) modes that lie within the
Reststrahlen band of AlN, which represent yet another avenue to
explore for multiband, frequency-tunable absorption in this system.
The coupling into SPhPs is weak in this particular parameter space;
this can be improved by a proper choice of geometric parameters,
which is outside the scope of the current work.

Nevertheless, our proof-of-concept perforated absorber shows a
measured absorptance of nearly 95\%,
Fig.~\ref{fig:optical}(e), higher than that of the flat absorber.
The measured flat and perforated cavities reach approximately 80\%
and 95\% peak absorptance, respectively. The nominal simulations, however,
predict near-unity absorptance for the flat film and 93\% for the
perforated film, the latter being the \(A_{\mathrm{cav}}\) used in
Section~IV. The origin of this
discrepancy is not entirely clear and requires further investigation,
although the different conditions in the release of flat versus perforated
films may be a contributing factor. The
LO and SPhP modes are also observed in measurements, as predicted by
simulations, although the SPhP mode is slightly blue-shifted compared
with the simulation, which may be due to fabrication imperfections
and nonlocal effects present in nanoscale phononic layers
\cite{gubbinOpticalNonlocalityPolar2020}. Remarkably, the extended
spectra show that no significant features appear below the main TO
band, all the way down to the mid-wave infrared (MWIR) range.

\section{Bolometric Measurement and Responsivity}

\begin{figure}[!t]
\centering
\includegraphics[width=\columnwidth]{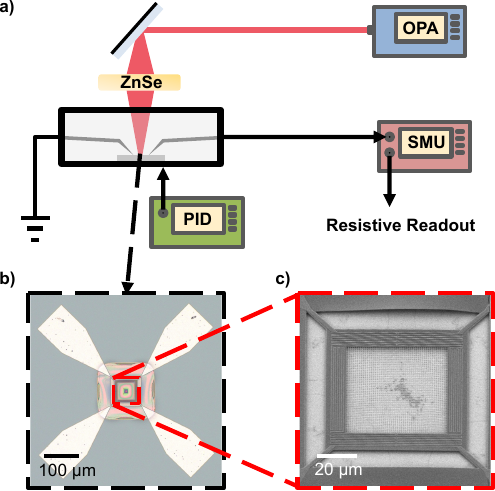}
\caption{
Infrared measurement setup and device images.
(a) Experimental setup for tunable infrared illumination and
resistance readout.
(b) Optical micrograph of the fabricated suspended device.
(c) SEM image of the central suspended absorber region.
}
\label{fig:setup}
\end{figure}

To validate these enhancements, the perforated AlN membranes were
integrated into a bolometric platform and characterized using the
setup in Fig.~\ref{fig:setup}(a). We employ an ORPHEUS-ONE tunable
optical parametric amplifier (OPA) with emission from
\SI{2}{\micro\meter} up to \SI{18}{\micro\meter} as the infrared
source. The suspended absorber is held in a vacuum chamber below
\SI{10}{\milli\torr}, attached to a PID-controlled stage that
maintains a constant temperature of \SI{293}{\kelvin} to avoid
thermal drift during testing. Under IR illumination through a ZnSe
lens, the device is probed for real-time electrical readout using a
source-measurement unit. Electrical readout is performed with a
Keithley 2450 sourcing \SI{0.3}{\milli\ampere} in a four-wire resistance
configuration. At each OPA center wavelength, resistance is recorded with
the IR source off and on after stabilization.
In other words, to quantify the device response, baseline resistance, IR OFF, is
subtracted from the active resistance, IR ON, to calculate the
absolute change in resistance:
\begin{equation}
\Delta R(\lambda)
=
R_{\mathrm{on}}(\lambda)-R_{\mathrm{off}}(\lambda).
\label{eq:deltaR}
\end{equation}
Hence, the wavelength dependence of this resistance change tracks
the spectral absorption of the suspended cavity and the perforated
AlN membrane, with the largest response occurring at wavelengths
where reflectance is minimized. The optical micrograph and SEM image
in Fig.~\ref{fig:setup}(b),(c) confirm successful fabrication of the
suspended membrane and the central perforated absorber region.

Because the suspended absorber is smaller than the incident optical
beam, the fraction of the beam power incident on the active
cavity footprint is calculated from the Gaussian beam profile.
We define the footprint \(\Omega\) as the full
\SI{50}{\micro\meter}\(\times\)\SI{50}{\micro\meter} suspended
cavity area, including the perforations. For
a Gaussian beam with total power \(P_0\) and \(1/e^2\) intensity
radius \(w\),
\begin{equation}
I(x,y)
=
\frac{2P_0}{\pi w^2}
\exp\left[
-\frac{2(x^2+y^2)}{w^2}
\right].
\label{eq:gaussian}
\end{equation}
The power incident on the cavity footprint is
\begin{equation}
P_{\mathrm{inc}}
=
\iint_{\Omega} I(x,y)\,dx\,dy,
\label{eq:footprint_power}
\end{equation}
with no perforation mask, because the cavity absorptance is
normalized to the full footprint. The
incident-power fraction is
\begin{equation}
\eta_{\mathrm{geo}}
=
\frac{\iint_{\Omega} I(x,y)\,dx\,dy}
{\iint_{-\infty}^{\infty} I(x,y)\,dx\,dy}.
\label{eq:power_fraction}
\end{equation}

\begin{figure}[!t]
\centering
\includegraphics[width=\columnwidth]{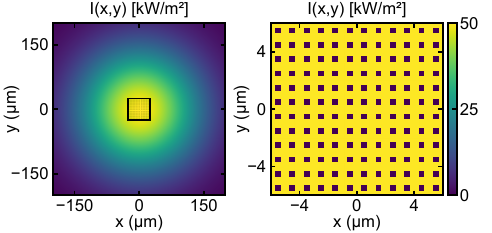}
\caption{
Gaussian beam-overlap calculation.
The left panel shows integration of the Gaussian beam profile over
the square absorber domain. The right panel shows a zoomed-in view
of the perforated absorber area. The full-footprint power overlap
is 3.90\%, whereas an AlN-only perforation mask gives 3.28\%.
}
\label{fig:beam_overlap}
\end{figure}

The beam profile and perforation mask are shown in
Fig.~\ref{fig:beam_overlap}. We perform a knife-edge measurement at
the sample plane to determine the beam radius \(w=\SI{200}{\micro\meter}\).
For this radius, Eq.~\eqref{eq:power_fraction} yields
\(\eta_{\mathrm{geo}}=0.0390\), the fraction of the beam power at the
sample plane that is incident on the full cavity footprint. Masking the perforations gives
3.28\%, the fraction incident on the remaining AlN. Note that this
smaller value is not multiplied by \(A_{\mathrm{cav}}\), since doing so
would account for the perforated area twice.

\begin{figure}[!t]
\centering
\includegraphics[width=\columnwidth]{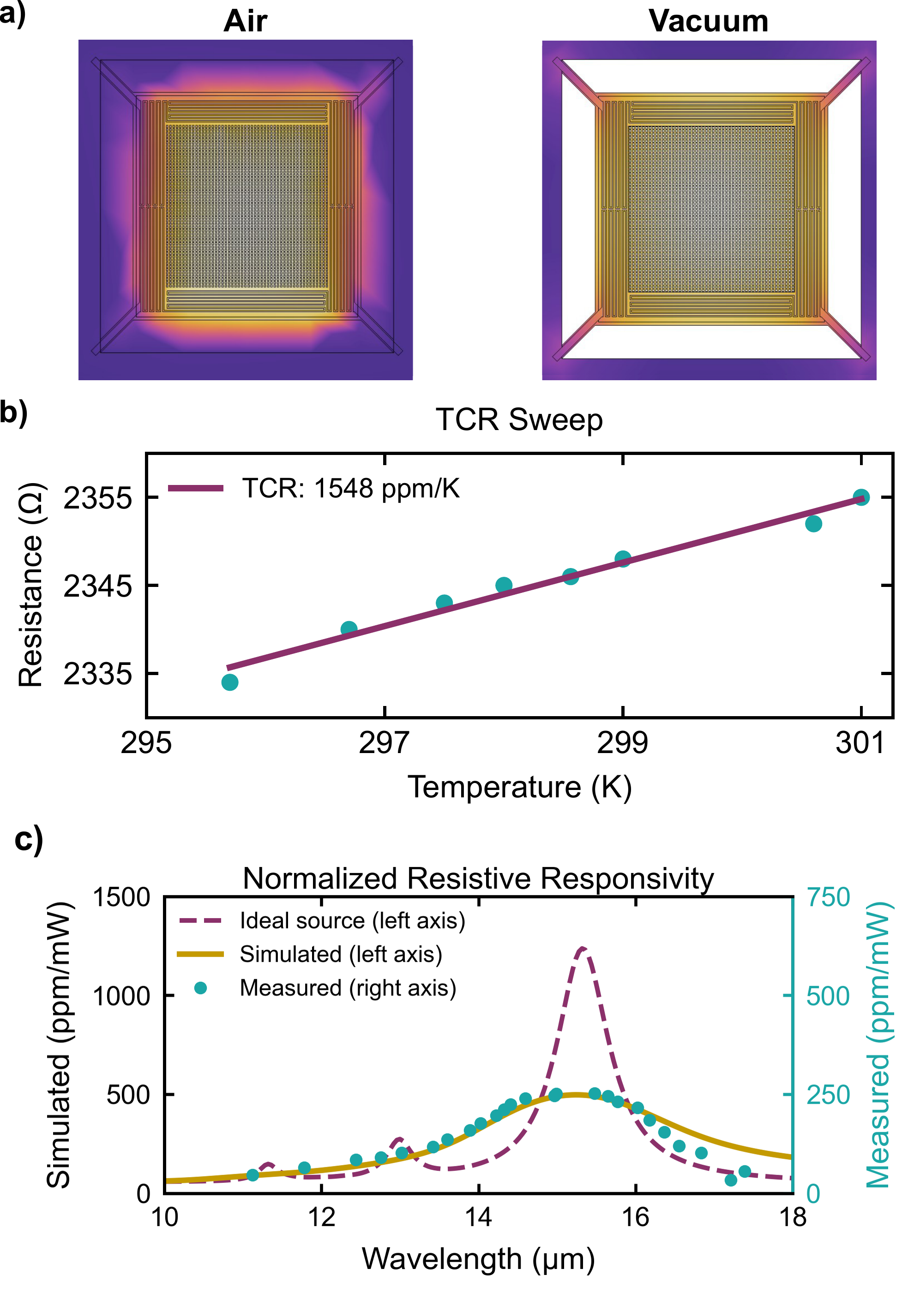}
\caption{
(a) Simulated temperature distribution of the suspended absorber
under infrared illumination in air and vacuum.
(b) Measured resistance as a function of temperature, yielding a
temperature coefficient of resistance of
\SI{1548}{\ppm\per\kelvin}.
(c) Simulated and measured spectral responsivity of the device.
\revThree{The simulated curves use the left axis, and the measured data use
the right axis, which is scaled to half of the left axis. The measured peak
(\SI{252}{\ppm\per\milli\watt}) is about half of the simulated OPA-source
peak (\SI{497}{\ppm\per\milli\watt}).}
}
\label{fig:responsivity}
\end{figure}

To validate the bolometric performance of our design, we demonstrate
that the measured resistive responsivity follows the wavelength dependence
expected from the cavity absorptance. All bolometric measurements are
performed below \SI{10}{\milli\torr}, measured at the pump. The experimental-vacuum model uses an effective air thermal conductivity of
\SI{1e-3}{\watt\per\meter\per\kelvin} \revThree{for all air in the model,
including the \SI{1}{\micro\meter} gap beneath the membrane,} and uses
\SI{25}{\watt\per\meter\per\kelvin} for the \SI{100}{\nano\meter} AlN film
\cite{wuCharacterizationThermalCrosstalk2009,perezHighThermalConductivity2023}.
Because both gas conduction and thin-film AlN conductivity depend on the
actual pressure, gap, and film properties, \(G_{\mathrm{th}}\) is treated as
a model estimate. This model gives
\begin{equation}
G_{\mathrm{th}}
=
\SI{1.58e-5}{\watt\per\kelvin},
\label{eq:Gth}
\end{equation}
which is used in the responsivity calculation. Atmospheric air gives
\SI{1.46e-4}{\watt\per\kelvin}, while zero gas conduction (vacuum) gives
\SI{4.85e-6}{\watt\per\kelvin}. The remaining zero-gas conductance and the
gradients near the anchors in Fig.~\ref{fig:responsivity}(a) show that the
tethers remain an alternative heat-loss path.

After determining the thermal conductance, the temperature
coefficient of resistance was experimentally extracted by measuring
the resistance shift of high-resistance meandering lines across a controlled temperature sweep, as shown in
Fig.~\ref{fig:responsivity}(b). The TCR is
\begin{equation}
\alpha
=
\frac{1}{R_0}\frac{dR}{dT},
\label{eq:TCR}
\end{equation}
and we obtain
\begin{equation}
\alpha
=
\SI{1548}{\ppm\per\kelvin}.
\label{eq:TCR_value}
\end{equation}

The measured normalized resistive responsivity is calculated as
\begin{equation}
\mathcal{R}_R(\lambda)
=
\frac{\Delta R(\lambda)}
{R_0P_{\mathrm{OPA}}(\lambda)}.
\label{eq:measured_responsivity}
\end{equation}
Here \(R_0\) is the IR-off baseline resistance at the applied
bias, and \(P_{\mathrm{OPA}}\) is the OPA output power measured before
the ZnSe lens and the 3~dB attenuator.
The absorbed power is modeled as
\begin{equation}
P_{\mathrm{abs}}(\lambda)
=
A_{\mathrm{cav}}(\lambda)
\eta_{\mathrm{geo}}
\eta_{\mathrm{opt}}
P_{\mathrm{OPA}}(\lambda),
\label{eq:absorbed_power}
\end{equation}
where \(A_{\mathrm{cav}}\) is the simulated absorptance of the complete
perforated cavity footprint and \(\eta_{\mathrm{opt}} = 0.35\) includes losses from the
focusing optics and attenuation in the optical path between the
power meter and the sample. The simulated
normalized resistive responsivity is therefore
\begin{equation}
\mathcal{R}_{R,\mathrm{sim}}(\lambda)
=
\frac{
\alpha
A_{\mathrm{cav}}(\lambda)
\eta_{\mathrm{geo}}
\eta_{\mathrm{opt}}
}{
G_{\mathrm{th}}
}.
\label{eq:expected_responsivity}
\end{equation}

Finally, we derive the simulated responsivity and compare it to the
measured results. We first consider an ideal source with a very
narrow linewidth below \SI{0.1}{\micro\meter}. Convolving this
idealized excitation with our simulated absorptance, we obtain a
spectral shape nearly identical to the initial simulation and FTIR
measurement. Our pulsed IR source, however, exhibits a broad spectral
profile that cannot fully resolve the narrowband resonances in our
device. Hence, convolving our source spectra with the absorptance
simulations yields a significantly broader, flatter response.

The effective cavity absorptance for an OPA spectrum centered at
\(\lambda_0\) is
\begin{equation}
A_{\mathrm{eff}}(\lambda_0)
=
\frac{
\int A_{\mathrm{cav}}(\lambda)S_{\lambda_0}(\lambda)\,d\lambda
}{
\int S_{\lambda_0}(\lambda)\,d\lambda
},
\label{eq:effective_absorptance}
\end{equation}
where \(S_{\lambda_0}(\lambda)\) is the measured OPA spectrum.
Supplementary Fig.~3 describes how \(S_{\lambda_0}\) is constructed
from the measured OPA spectra for Fig.~\ref{fig:responsivity}(c).
The source-broadened simulated responsivity is therefore
\begin{equation}
\mathcal{R}_{R,\mathrm{OPA}}(\lambda_0)
=
\frac{
\alpha
A_{\mathrm{eff}}(\lambda_0)
\eta_{\mathrm{geo}}
\eta_{\mathrm{opt}}
}{
G_{\mathrm{th}}
}.
\label{eq:opa_responsivity}
\end{equation}

For \(\alpha=\SI{1548}{\ppm\per\kelvin}\),
\(\eta_{\mathrm{geo}}=0.0390\), \(\eta_{\mathrm{opt}}=0.35\), and
\(G_{\mathrm{th}}=\SI{1.58e-5}{\watt\per\kelvin}\), the maximum
responsivity for unity absorptance is approximately
\SI{1337}{\ppm\per\milli\watt}. The simulated responsivity peaks at
\SI{1237}{\ppm\per\milli\watt} for the ideal source and at
\SI{497}{\ppm\per\milli\watt} for the OPA source.
As shown in Fig.~\ref{fig:responsivity}(c), the latter
follows the spectral shape of our measured results,
although the measured peak is about half as large. This discrepancy is likely the result of multiple deviations in our modeling, such as thermal conductivity variance in the thin sputtered film and vacuum chamber air.
The \SI{0.772}{\micro\meter} FWHM
reported for this work is extracted from the FTIR absorptance
spectrum of the cavity. The OPA measurement verifies bolometric response near the same
resonance, but its measured width is broadened by the OPA linewidth. We
demonstrate a peak normalized resistive responsivity of
\SI{252}{\ppm\per\milli\watt} at \SI{15.48}{\micro\meter}.
To our knowledge, this is the first thin-film microbolometer to
demonstrate narrowband bolometric response in the \SIrange{14}{18}{\micro\meter}
range. The resonance near \SI{15.4}{\micro\meter} overlaps a strong CO$_2$
absorption band, suggesting potential use in CO$_2$-sensitive spectroscopy
\cite{
aumannApplicationAtmosphericInfrared2004,
ziebaNoThickCarbon2023
}.
Table~\ref{tab:comparison} compares spectral absorber
characteristics only. Responsivity, noise, response time, and operating
conditions differ among the cited devices and are not directly compared.

\begin{table}[!t]
\centering
\caption{Comparison to CMOS-Compatible Infrared Detectors}
\label{tab:comparison}
\small
\renewcommand{\arraystretch}{1.08}

\begin{tabular*}{\columnwidth}
{@{\extracolsep{\fill}}lcccc@{}}
\toprule
Work &
\makecell{Range\\($\mu$m)} &
\makecell{FWHM\\($\mu$m)} &
\makecell{Peak\\Abs.} &
\makecell{Area\\($\mu$m$^2$)} \\
\midrule

Goldsmith \textit{et al.}
\cite{goldsmithLongwaveInfraredSelective2017a}
& 8--9
& 0.332
& 0.79
& $>10^3$ \\

G\"ulseren \textit{et al.}
\cite{gulserenExperimentalStudySpectrally2024}
& 3.2--6.1
& 0.450
& 0.95
& 7500 \\

Hui \textit{et al.}
\cite{huiPlasmonicPiezoelectricNanomechanical2016}
& 8--10
& 1.50
& 0.80
& 15000 \\

Doan \textit{et al.}
\cite{doanMEMSBasedQuadWavelength2019}
& 3.3--4.5
& 0.35
& 0.92
& 40000 \\

This work
& 15--16
& 0.772
& 0.95
& 2500 \\

\bottomrule
\end{tabular*}

\end{table}

\section{Conclusion}

In conclusion, we have demonstrated a thin-film microbolometer
architecture for narrowband infrared detection in the VLWIR range.
The device employs a \SI{100}{\nano\meter} suspended AlN membrane
above a Pt back reflector, leveraging the AlN transverse optical
phonon for spectrally selective cavity absorption near
\SI{15.4}{\micro\meter} and high out-of-band rejection. FTIR reflection
measurements performed directly on the released bolometer show nearly
95\% peak cavity absorptance with an intrinsic absorptance FWHM of
\SI{0.772}{\micro\meter}. Bolometric operation is confirmed via DC
resistance measurements under tunable infrared illumination, with a
measured TCR of \SI{1548}{\ppm\per\kelvin} and a peak normalized
resistive responsivity of \SI{252}{\ppm\per\milli\watt} at
\SI{15.48}{\micro\meter}. These results establish
suspended thin-film AlN as a viable platform for compact,
wavelength-selective thermal detectors in the VLWIR. Future work includes geometric tuning
of the perforation array to access surface-phonon-polariton
resonances within the AlN Reststrahlen band, as well as optimizing
thermal conductance and bias current to improve responsivity.





\balance

\bibliographystyle{IEEEtran}
\bibliography{references}

@book{novotny_principles_2012,
  title = {Principles of {Nano-Optics}},
  author = {Novotny, Lukas and Hecht, Bert},
  year = {2012},
  edition = {2nd},
  address = {Cambridge, U.K.},
  publisher = {Cambridge University Press},
  googlebooks = {RHC\_AwAAQBAJ},
  isbn = {978-1-139-56045-0},
  langid = {english}
}

@article{abdullahUncooledTwomicrobolometerStack2023,
  title = {Uncooled Two-Microbolometer Stack for Long Wavelength Infrared Detection},
  author = {Abdullah, Amjed and Koppula, Akshay and Alkorjia, Omar and Almasri, Mahmoud},
  year = 2023,
  month = mar,
  journal = {Sci. Rep.},
  volume = {13},
  number = {1},
  pages = {3470},
  publisher = {Nature Publishing Group},
  issn = {2045-2322},
  doi = {10.1038/s41598-023-30328-1},
  urldate = {2026-03-17},
  copyright = {2023 The Author(s)},
  langid = {english}
}

@article{alkorjiaMetasurfaceEnabledUncooled2023,
  title = {Metasurface-{{Enabled Uncooled Silicon Germanium Oxide Infrared Microbolometers}} for {{Long-Wave Infrared Detection}}},
  author = {Alkorjia, Omar and Koppula, Akshay and Abdullah, Amjed and Almasri, Mahmoud},
  year = 2023,
  month = aug,
  journal = {IEEE Sensors J.},
  volume = {23},
  number = {15},
  pages = {16616--16624},
  doi = {10.1109/JSEN.2023.3282173}
}

@article{beliaev2021thickness,
  title = {Thickness-Dependent Optical Properties of Aluminum Nitride Films for Mid-Infrared Wavelengths},
  author = {Beliaev, Leonid Yu. and Shkondin, Evgeniy and Lavrinenko, Andrei V. and Takayama, Osamu},
  year = 2021,
  month = jul,
  journal = {J. Vac. Sci. Technol. A},
  volume = {39},
  number = {4},
  pages = {043408},
  issn = {0734-2101, 1520-8559},
  doi = {10.1116/6.0000884},
  urldate = {2026-04-11},
  langid = {english}
}

@article{caldwellLowlossInfraredTerahertz2014a,
  title = {Low-loss, Infrared and Terahertz Nanophotonics Using Surface Phonon Polaritons},
  author = {Caldwell, Joshua D. and Lindsay, Lucas and Giannini, Vincenzo and Vurgaftman, Igor and Reinecke, Thomas L. and Maier, Stefan A. and Glembocki, Orest J.},
  year = 2015,
  journal = {Nanophotonics},
  volume = {4},
  number = {1},
  pages = {44--68},
  issn = {2192-8614, 2192-8614},
  doi = {10.1515/nanoph-2014-0003},
  urldate = {2026-04-11},
  copyright = {http://creativecommons.org/licenses/by-nc-nd/3.0/},
  langid = {english}
}

@article{chenEnhancementCAxisOriented2021,
  title = {Enhancement of C-{{Axis Oriented Aluminum Nitride Films}} via {{Low Temperature DC Sputtering}}},
  author = {Chen, Ze-Hui and Li, Chengying and Chen, Yueh-Han and Chu, Shengyuan and Tsai, Cheng-Che and Hong, Cheng-Shong},
  year = 2021,
  month = aug,
  journal = {IEEE Sensors J.},
  volume = {21},
  number = {16},
  pages = {17673--17677},
  issn = {1558-1748},
  doi = {10.1109/JSEN.2021.3077274},
  urldate = {2026-03-19}
}

@inproceedings{felmetsgerCrystalOrientationStress2008,
  title = {Crystal Orientation and Stress in {{AC}} Reactively Sputtered {{AlN}} Films on {{Mo}} Electrodes for Electro-Acoustic Devices},
  booktitle = {2008 {{IEEE Ultrasonics Symposium}}},
  author = {Felmetsger, Valery V. and Laptev, Pavel N. and Tanner, Shawn M.},
  year = 2008,
  month = nov,
  pages = {2146--2149},
  publisher = {IEEE},
  address = {Beijing, China},
  doi = {10.1109/ULTSYM.2008.0531},
  urldate = {2026-03-20},
  isbn = {978-1-4244-2428-3},
  langid = {english}
}

@article{foteinopoulouPhononpolaritonicsEnablingPowerful2018,
  title = {Phonon-polaritonics: Enabling Powerful Capabilities for Infrared Photonics},
  shorttitle = {Phonon-polaritonics},
  author = {Foteinopoulou, Stavroula and Devarapu, Ganga Chinna Rao and Subramania, Ganapathi S. and Krishna, Sanjay and Wasserman, Daniel},
  year = 2019,
  journal = {Nanophotonics},
  volume = {8},
  number = {12},
  pages = {2129--2175},
  issn = {2192-8614, 2192-8614},
  doi = {10.1515/nanoph-2019-0232},
  urldate = {2026-04-11},
  copyright = {http://creativecommons.org/licenses/by/4.0/},
  langid = {english}
}

@article{goldsmithLongwaveInfraredSelective2017a,
  title = {Long-Wave Infrared Selective Pyroelectric Detector Using Plasmonic near-Perfect Absorbers and Highly Oriented Aluminum Nitride},
  author = {Goldsmith, John H. and Vangala, Shivashankar and Hendrickson, Joshua R. and Cleary, Justin W. and Vella, Jarrett H.},
  year = 2017,
  month = sep,
  journal = {J. Opt. Soc. Am. B},
  volume = {34},
  number = {9},
  pages = {1965},
  issn = {0740-3224, 1520-8540},
  doi = {10.1364/JOSAB.34.001965},
  urldate = {2026-04-13},
  langid = {english}
}

@article{gubbinOpticalNonlocalityPolar2020,
  title = {Optical {{Nonlocality}} in {{Polar Dielectrics}}},
  author = {Gubbin, Christopher R. and De Liberato, Simone},
  year = 2020,
  month = may,
  journal = {Phys. Rev. X},
  volume = {10},
  number = {2},
  pages = {021027},
  issn = {2160-3308},
  doi = {10.1103/PhysRevX.10.021027},
  urldate = {2026-04-11},
  langid = {english}
}

@article{gubbinSurfacePhononPolaritons2022,
  title = {Surface Phonon Polaritons for Infrared Optoelectronics},
  author = {Gubbin, Christopher R. and De Liberato, Simone and Folland, Thomas G.},
  year = 2022,
  month = jan,
  journal = {J. Appl. Phys.},
  volume = {131},
  number = {3},
  pages = {030901},
  issn = {0021-8979, 1089-7550},
  doi = {10.1063/5.0064234},
  urldate = {2026-04-11},
  langid = {english}
}

@article{iqbalReactiveSputteringAluminum2018,
  title = {Reactive {{Sputtering}} of {{Aluminum Nitride}} (002) {{Thin Films}} for {{Piezoelectric Applications}}: {{A Review}}},
  shorttitle = {Reactive {{Sputtering}} of {{Aluminum Nitride}} (002) {{Thin Films}} for {{Piezoelectric Applications}}},
  author = {Iqbal, Abid and {Mohd-Yasin}, Faisal},
  year = 2018,
  month = jun,
  journal = {Sensors},
  volume = {18},
  number = {6},
  pages = {1797},
  publisher = {Multidisciplinary Digital Publishing Institute},
  issn = {1424-8220},
  doi = {10.3390/s18061797},
  urldate = {2026-03-20},
  copyright = {http://creativecommons.org/licenses/by/3.0/},
  langid = {english}
}

@article{jungWavelengthSelectiveInfraredMetasurface2015,
  title = {Wavelength-{{Selective Infrared Metasurface Absorber}} for {{Multispectral Thermal Detection}}},
  author = {Jung, Joo-Yun and Lee, Jihye and Choi, Dae-Geun and Choi, Jun-Hyuk and Jeong, Jun-Ho and Lee, Eung-Sug and Neikirk, Dean P.},
  year = 2015,
  month = dec,
  journal = {IEEE Photonics J.},
  volume = {7},
  number = {6},
  pages = {6804210},
  issn = {1943-0655},
  doi = {10.1109/JPHOT.2015.2504975},
  urldate = {2026-04-11},
  copyright = {https://ieeexplore.ieee.org/Xplorehelp/downloads/license-information/OAPA.html}
}

@article{kozuchSurfaceenhancedInfraredAbsorption2023,
  title = {Surface-Enhanced Infrared Absorption Spectroscopy},
  author = {Kozuch, Jacek and Ataka, Kenichi and Heberle, Joachim},
  year = 2023,
  month = sep,
  journal = {Nat. Rev. Methods Primers},
  volume = {3},
  number = {1},
  pages = {70},
  publisher = {Nature Publishing Group},
  issn = {2662-8449},
  doi = {10.1038/s43586-023-00253-8},
  urldate = {2026-03-17},
  copyright = {2023 Springer Nature Limited},
  langid = {english}
}

@article{liuPiezoelectricThinFilms2025,
  title = {Piezoelectric Thin Films and Their Applications in {{MEMS}}: {{A}} Review},
  shorttitle = {Piezoelectric Thin Films and Their Applications in {{MEMS}}},
  author = {Liu, Jinpeng and Tan, Hua and Zhou, Xinyi and Ma, Weigang and Wang, Chuanmin and Tran, Nguyen-Minh-An and Lu, Wenlong and Chen, Feng and Wang, Junya and Zhang, Haibo},
  year = 2025,
  month = jan,
  journal = {J. Appl. Phys.},
  volume = {137},
  number = {2},
  pages = {020702},
  issn = {0021-8979},
  doi = {10.1063/5.0244749},
  urldate = {2026-03-21}
}

@article{huoBroadbandAbsorptionBased2021,
  title = {Broadband {{Absorption Based}} on {{Thin Refractory Titanium Nitride Patterned Film Metasurface}}},
  author = {Huo, Dewang and Ma, Xinyan and Su, Hang and Wang, Chao and Zhao, Hua},
  year = 2021,
  month = apr,
  journal = {Nanomaterials},
  volume = {11},
  number = {5},
  pages = {1092},
  issn = {2079-4991},
  doi = {10.3390/nano11051092},
  urldate = {2026-04-14},
  langid = {english}
}

@article{luhmannUltrathin2Nm2020,
  title = {Ultrathin 2 Nm Gold as Impedance-Matched Absorber for Infrared Light},
  author = {Luhmann, Niklas and H{\o}j, Dennis and Piller, Markus and K{\"a}hler, Hendrik and Chien, Miao-Hsuan and West, Robert G. and Andersen, Ulrik Lund and Schmid, Silvan},
  year = 2020,
  month = may,
  journal = {Nat. Commun.},
  volume = {11},
  number = {1},
  pages = {2161},
  issn = {2041-1723},
  doi = {10.1038/s41467-020-15762-3},
  urldate = {2026-04-11},
  langid = {english}
}

@article{lundhResidualStressAnalysis2021,
  title = {Residual Stress Analysis of Aluminum Nitride Piezoelectric Micromachined Ultrasonic Transducers Using {{Raman}} Spectroscopy},
  author = {Lundh, James Spencer and Coleman, Kathleen and Song, Yiwen and Griffin, Benjamin A. and Esteves, Giovanni and Douglas, Erica A. and Edstrand, Adam and Badescu, Stefan C. and Moore, Elizabeth A. and Leach, Jacob H. and Moody, Baxter and {Trolier-McKinstry}, Susan and Choi, Sukwon},
  year = 2021,
  month = jul,
  journal = {J. Appl. Phys.},
  volume = {130},
  number = {4},
  pages = {044501},
  issn = {0021-8979, 1089-7550},
  doi = {10.1063/5.0056302},
  urldate = {2026-03-19},
  langid = {english}
}

@article{martiniUncooledThermalInfrared2025,
  title = {Uncooled Thermal Infrared Detection near the Fundamental Limit Using a Silicon Nitride Nanomechanical Resonator with a Broadband Absorber},
  author = {Martini, Paolo and Kanellopulos, Kostas and Emminger, Stefan and Luhmann, Niklas and Piller, Markus and West, Robert G. and Schmid, Silvan},
  year = 2025,
  month = apr,
  journal = {Commun. Phys.},
  volume = {8},
  number = {1},
  pages = {166},
  issn = {2399-3650},
  doi = {10.1038/s42005-025-02093-2},
  urldate = {2026-04-11},
  langid = {english}
}

@article{nordinMidinfraredEpsilonnearzeroModes2017,
  title = {Mid-Infrared Epsilon-near-Zero Modes in Ultra-Thin Phononic Films},
  author = {Nordin, L. and Dominguez, O. and Roberts, C. M. and Streyer, W. and Feng, K. and Fang, Z. and Podolskiy, V. A. and Hoffman, A. J. and Wasserman, D.},
  year = 2017,
  month = aug,
  journal = {Appl. Phys. Lett.},
  volume = {111},
  number = {9},
  pages = {091105},
  issn = {0003-6951, 1077-3118},
  doi = {10.1063/1.4996213},
  urldate = {2026-04-11},
  langid = {english}
}

@article{perezHighThermalConductivity2023,
  title = {High {{Thermal Conductivity}} of {{Submicrometer Aluminum Nitride Thin Films Sputter-Deposited}} at {{Low Temperature}}},
  author = {Perez, Christopher and McLeod, Aaron J. and Chen, Michelle E. and Yi, Su-in and Vaziri, Sam and Hood, Ryan and Ueda, Scott T. and Bao, Xinyu and Asheghi, Mehdi and Park, Woosung and Talin, A. Alec and Kumar, Suhas and Pop, Eric and Kummel, Andrew C. and Goodson, Kenneth E.},
  year = 2023,
  month = nov,
  journal = {ACS Nano},
  volume = {17},
  number = {21},
  pages = {21240--21250},
  issn = {1936-0851, 1936-086X},
  doi = {10.1021/acsnano.3c05485},
  urldate = {2026-04-11},
  copyright = {https://doi.org/10.15223/policy-029},
  langid = {english}
}

@article{sakoticMidInfraredPerfectAbsorption2024,
  title = {Mid-{{Infrared Perfect Absorption}} with {{Planar}} and {{Subwavelength-Perforated Ultrathin Metal Films}}},
  author = {Sakotic, Zarko and Raju, Amogh and Ware, Alexander and Est{\'e}vez H., F{\'e}lix A. and Brown, Madeline and Magendzo Behar, Yonathan and Hungund, Divya and Wasserman, Daniel},
  year = 2024,
  journal = {Adv. Phys. Res.},
  volume = {3},
  number = {8},
  pages = {2400012},
  issn = {2751-1200},
  doi = {10.1002/apxr.202400012},
  urldate = {2026-02-05},
  langid = {english}
}

@article{sakoticPerfectAbsorptionUltimate2023,
  title = {Perfect {{Absorption}} at the {{Ultimate Thickness Limit}} in {{Planar Films}}},
  author = {Sakotic, Zarko and Ware, Alexander and Povinelli, Michelle and Wasserman, Daniel},
  year = 2023,
  month = dec,
  journal = {ACS Photonics},
  volume = {10},
  number = {12},
  pages = {4244--4251},
  issn = {2330-4022, 2330-4022},
  doi = {10.1021/acsphotonics.3c01017},
  urldate = {2026-04-11},
  copyright = {https://doi.org/10.15223/policy-029},
  langid = {english}
}

@article{saleemInfraredPhotodetectorsRecent2024,
  title = {Infrared {{Photodetectors}}: {{Recent Advances}} and {{Challenges Toward Innovation}} for {{Image Sensing Applications}}},
  author = {Saleem, Muhammad Imran and Kyaw, Aung Ko Ko and Hur, Jaehyun},
  year = 2024,
  month = nov,
  journal = {Adv. Opt. Mater.},
  volume = {12},
  number = {33},
  pages = {2401625},
  doi = {10.1002/adom.202401625},
  langid = {english}
}

@article{talghaderSpectralSelectivityInfrared2012,
  title = {Spectral Selectivity in Infrared Thermal Detection},
  author = {Talghader, Joseph J. and Gawarikar, Anand S. and Shea, Ryan P.},
  year = 2012,
  month = aug,
  journal = {Light Sci. Appl.},
  volume = {1},
  number = {8},
  pages = {e24},
  publisher = {Nature Publishing Group},
  issn = {2047-7538},
  doi = {10.1038/lsa.2012.24},
  urldate = {2026-03-17},
  copyright = {2012 The Author(s)},
  langid = {english}
}

@article{tremlHighResolutionDetermination2016,
  title = {High Resolution Determination of Local Residual Stress Gradients in Single- and Multilayer Thin Film Systems},
  author = {Treml, R. and Kozic, D. and Zechner, J. and Maeder, X. and Sartory, B. and G{\"a}nser, H.-P. and Sch{\"o}ngrundner, R. and Michler, J. and Brunner, R. and Kiener, D.},
  year = 2016,
  month = jan,
  journal = {Acta Mater.},
  volume = {103},
  pages = {616--623},
  issn = {13596454},
  doi = {10.1016/j.actamat.2015.10.044},
  urldate = {2026-03-27},
  langid = {english}
}

@article{wangHighspeedLongwaveInfrared2024,
  title = {High-Speed Long-Wave Infrared Ultra-Thin Photodetectors},
  author = {Wang, Yinan and Muhowski, Aaron J. and Nordin, Leland and Dev, Sukrith and Allen, Monica and Allen, Jeffery and Wasserman, Daniel},
  year = 2024,
  month = jan,
  journal = {APL Photonics},
  volume = {9},
  number = {1},
  pages = {016117},
  issn = {2378-0967},
  doi = {10.1063/5.0181052},
  urldate = {2026-04-11},
  langid = {english}
}

@article{wareDecouplingAbsorptionRadiative2023,
  title = {Decoupling Absorption and Radiative Cooling in Mid-Wave Infrared Bolometric Elements},
  author = {Ware, Alexander and Bergthold, Morgan and Mansfield, Noah and Sakotic, Zarko and Scott, Ethan A. and Harris, C. Thomas and Wasserman, Daniel},
  year = 2023,
  month = jun,
  journal = {Opt. Lett.},
  volume = {48},
  number = {12},
  pages = {3155},
  issn = {0146-9592, 1539-4794},
  doi = {10.1364/OL.491601},
  urldate = {2026-04-11},
  langid = {english}
}

@article{woltersdorffBerOptischenKonstanten1934,
  title = {{\"Uber die optischen Konstanten d\"unner Metallschichten im langwelligen Ultrarot}},
  author = {Woltersdorff, Wilhelm},
  year = 1934,
  month = mar,
  journal = {Z. Physik},
  volume = {91},
  number = {3-4},
  pages = {230--252},
  issn = {1434-6001, 1434-601X},
  doi = {10.1007/BF01341647},
  urldate = {2026-04-11},
  copyright = {http://www.springer.com/tdm},
  langid = {german}
}

@article{zhaoUltrathinMXeneAssemblies2023,
  title = {Ultrathin {{MXene}} Assemblies Approach the Intrinsic Absorption Limit in the 0.5--10 {{THz}} Band},
  author = {Zhao, Tao and Xie, Peiyao and Wan, Hujie and Ding, Tianpeng and Liu, Mengqi and Xie, Jinlin and Li, Enen and Chen, Xuequan and Wang, Tianwu and Zhang, Qing and Wei, Yanyu and Gong, Yubin and Wen, Qiye and Hu, Min and Qiu, Cheng-Wei and Xiao, Xu},
  year = 2023,
  month = jul,
  journal = {Nat. Photon.},
  volume = {17},
  number = {7},
  pages = {622--628},
  issn = {1749-4885, 1749-4893},
  doi = {10.1038/s41566-023-01197-x},
  urldate = {2026-04-11},
  langid = {english}
}

@article{nordinUltrathinPlasmonicDetectors2021,
  title = {Ultra-Thin Plasmonic Detectors},
  author = {Nordin, Leland and Petluru, Priyanka and Kamboj, Abhilasha and Muhowski, Aaron J. and Wasserman, Daniel},
  year = 2021,
  month = dec,
  journal = {Optica},
  volume = {8},
  number = {12},
  pages = {1545},
  issn = {2334-2536},
  doi = {10.1364/OPTICA.438039},
  urldate = {2026-04-18},
  langid = {english}
}

@inproceedings{aumannApplicationAtmosphericInfrared2004,
  title = {Application of {{Atmospheric Infrared Sounder}} ({{AIRS}}) Data to Climate Research},
  booktitle = {Proc. {SPIE} 5570, Sensors, Systems, and Next-Generation Satellites {VIII}},
  author = {Aumann, Hartmut H. and Gregorich, David T. and Gaiser, Steve L. and Chahine, Moustafa T.},
  editor = {Meynart, Roland and Neeck, Steven P. and Shimoda, Haruhisa},
  year = 2004,
  month = nov,
  pages = {202},
  address = {Maspalomas, Canary Islands, Spain},
  doi = {10.1117/12.565712},
  urldate = {2026-04-18}
}

@article{ziebaNoThickCarbon2023,
  title = {No Thick Carbon Dioxide Atmosphere on the Rocky Exoplanet {{TRAPPIST-1}} c},
  author = {Zieba, Sebastian and Kreidberg, Laura and Ducrot, Elsa and Gillon, Micha{\"e}l and Morley, Caroline and Schaefer, Laura and Tamburo, Patrick and Koll, Daniel D. B. and Lyu, Xintong and Acu{\~n}a, Lorena and Agol, Eric and Iyer, Aishwarya R. and Hu, Renyu and Lincowski, Andrew P. and Meadows, Victoria S. and Selsis, Franck and Bolmont, Emeline and Mandell, Avi M. and Suissa, Gabrielle},
  year = 2023,
  month = aug,
  journal = {Nature},
  volume = {620},
  number = {7975},
  pages = {746--749},
  issn = {0028-0836, 1476-4687},
  doi = {10.1038/s41586-023-06232-z},
  urldate = {2026-04-18},
  langid = {english}
}

@article{abedinidereshgiLithographyfreeIRPolarization2020,
  title = {Lithography-Free {{IR}} Polarization Converters via Orthogonal in-Plane Phonons in {$\alpha$}-{{MoO3}} Flakes},
  author = {Abedini Dereshgi, Sina and Folland, Thomas G. and Murthy, Akshay A. and Song, Xianglian and Tanriover, Ibrahim and Dravid, Vinayak P. and Caldwell, Joshua D. and Aydin, Koray},
  year = 2020,
  month = nov,
  journal = {Nat. Commun.},
  volume = {11},
  number = {1},
  pages = {5771},
  issn = {2041-1723},
  doi = {10.1038/s41467-020-19499-x},
  urldate = {2026-04-18},
  langid = {english}
}

@article{lingNearUnityLightMatter2024,
  title = {Near-{{Unity Light}}--{{Matter Interaction}} in {{Mid-Infrared}} van Der {{Waals Metasurfaces}}},
  author = {Ling, Haonan and Nourbakhsh, Milad and Whiteside, Vincent R. and Tischler, Joseph G. and Davoyan, Artur R.},
  year = 2024,
  month = mar,
  journal = {Nano Lett.},
  volume = {24},
  number = {11},
  pages = {3315--3322},
  issn = {1530-6984, 1530-6992},
  doi = {10.1021/acs.nanolett.3c04118},
  urldate = {2026-04-18},
  copyright = {https://doi.org/10.15223/policy-029},
  langid = {english}
}

@article{mastrangeloAdhesionrelatedFailureMechanisms1997,
  title = {Adhesion-Related Failure Mechanisms in Micromechanical Devices},
  author = {Mastrangelo, C. H.},
  year = 1997,
  month = sep,
  journal = {Tribol. Lett.},
  volume = {3},
  number = {3},
  pages = {223--238},
  issn = {1023-8883, 1573-2711},
  doi = {10.1023/A:1019133222401},
  urldate = {2026-04-20},
  copyright = {https://www.springernature.com/gp/researchers/text-and-data-mining},
  langid = {english}
}

@article{wuCharacterizationThermalCrosstalk2009,
  author  = {Wu, Huilai and Grabarnik, Semen and Emadi, A. and de Graaf, G. and Wolffenbuttel, R. F.},
  title   = {Characterization of thermal cross-talk in a {MEMS}-based thermopile detector array},
  journal = {J. Micromech. Microeng.},
  volume  = {19},
  number  = {7},
  pages   = {074022},
  year    = {2009},
  doi     = {10.1088/0960-1317/19/7/074022}
}

@article{gulserenExperimentalStudySpectrally2024,
  author  = {G{\"u}lseren, Melisa Ekin and Benson, Matthew and Parker, Ryan W. and Segovia-Fernandez, Jeronimo and Yen, Ernest Ting-Ta and G{\'o}mez-D{\'i}az, J. Sebasti{\'a}n},
  title   = {Experimental Study of Spectrally Selective {MEMS}/Metasurface Infrared Detectors},
  journal = {IEEE Sensors J.},
  volume  = {24},
  number  = {11},
  pages   = {17313--17323},
  year    = {2024},
  doi     = {10.1109/JSEN.2024.3388966}
}

@article{huiPlasmonicPiezoelectricNanomechanical2016,
  author  = {Hui, Yu and Gomez-Diaz, Juan Sebastian and Qian, Zhenyun and Al{\`u}, Andrea and Rinaldi, Matteo},
  title   = {Plasmonic piezoelectric nanomechanical resonator for spectrally selective infrared sensing},
  journal = {Nat. Commun.},
  volume  = {7},
  pages   = {11249},
  year    = {2016},
  doi     = {10.1038/ncomms11249}
}

@article{doanMEMSBasedQuadWavelength2019,
  author  = {Doan, Anh Tung and Yokoyama, Takahiro and Dao, Thang Duy and Ishii, Satoshi and Ohi, Akihiko and Nabatame, Toshihide and Wada, Yoshiki and Maruyama, Shigenao and Nagao, Tadaaki},
  title   = {A {MEMS}-Based Quad-Wavelength Hybrid Plasmonic--Pyroelectric Infrared Detector},
  journal = {Micromachines},
  volume  = {10},
  number  = {6},
  pages   = {413},
  year    = {2019},
  doi     = {10.3390/mi10060413}
}

\end{document}